\documentclass[sigconf]{acmart}
\usepackage{soul}
\usepackage{rotating}
\usepackage{pifont}
\usepackage[shortlabels]{enumitem}
\usepackage{fontawesome5}
\usepackage{subcaption}

\newcommand*\circled[1]
{\ifnum#1<1 %
    \@ctrerr
  \else
    \ifnum#1>10 %
      \@ctrerr
    \else
      \scalebox{1.1}{\ding{\the\numexpr 201+(#1)\relax}}%
    \fi
  \fi
}

\copyrightyear{2024}
\acmYear{2024}
\setcopyright{acmlicensed}\acmConference[DIS '24]{Designing Interactive Systems Conference}{July 1--5, 2024}{IT University of Copenhagen, Denmark}
\acmBooktitle{Designing Interactive Systems Conference (DIS '24), July 1--5, 2024, IT University of Copenhagen, Denmark}
\acmDOI{10.1145/3643834.3661544}
\acmISBN{979-8-4007-0583-0/24/07}

\begin{document}

\title[Motivating Users to Attend to Privacy: A Theory-Driven Design Study]{Motivating Users to Attend to Privacy:\\A Theory-Driven Design Study}

\author{Varun Shiri}
\authornote{The first two authors contributed equally to the paper.}
\email{varun.shiri@polymtl.ca}
\affiliation{%
  \department{Department of Computer and Software Engineering}
  \institution{Polytechnique Montreal}
  \city{Montreal}
  \state{QC}
  \country{Canada}
}
\author{Maggie Xiong}
\authornotemark[1]
\email{maggie.xiong@mail.mcgill.ca}
\affiliation{%
  \department{School of Computer Science}
  \institution{McGill University}
  \city{Montreal}
  \state{QC}
  \country{Canada}
}
\author{Jinghui Cheng}
\email{jinghui.cheng@polymtl.ca}
\affiliation{%
  \department{Department of Computer and Software Engineering}
  \institution{Polytechnique Montreal}
  \city{Montreal}
  \state{QC}
  \country{Canada}
}
\author{Jin L.C. Guo}
\email{jin.guo@mcgill.ca}
\affiliation{%
  \department{School of Computer Science}
  \institution{McGill University}
  \city{Montreal}
  \state{QC}
  \country{Canada}
}

\begin{abstract}
In modern technology environments, raising users' privacy awareness is crucial. Existing efforts largely focused on privacy policy presentation and failed to systematically address a radical challenge of user motivation for initiating privacy awareness. Leveraging the Protection Motivation Theory (PMT), we proposed design ideas and categories dedicated to motivating users to engage with privacy-related information. Using these design ideas, we created a conceptual prototype, enhancing the current App Store product page. Results from an online experiment and follow-up interviews showed that our design effectively motivated participants to attend to privacy issues, raising both the threat appraisal and coping appraisal, two main factors in PMT. Our work indicated that effective design should consider combining PMT components, calibrating information content, and integrating other design elements, such as visual cues and user familiarity. Overall, our study contributes valuable design considerations driven by the PMT to amplify the motivational aspect of privacy communication.
\end{abstract}




\maketitle

\section{Introduction}

In today's technological landscape, every user inevitably encounters a plethora of applications and services that produce, collect, analyze, and disseminate their personal data. While some of these practices can be useful, aiming at enhanced user experiences, others do not directly benefit users, gearing towards profiting from targeted advertisements or other financial avenues. Regardless of the intention of the app and service providers, users should be provided with opportunities to understand and have control over the ways their personal data is handled by those applications and services. Over time, many regulations and guidelines have been implemented, such as the Privacy Act in Canada~\cite{CanadaPrivacyAct}, the OECD guidelines~\cite{OECDGuidelines}, and more recently, GPDR~\cite{gdpr}. These regulations and guidelines predominantly focus on regulating the conduct of companies and developers, including compelling them to communicate their privacy practices to users primarily through privacy policies. As such, the privacy policy has become an essential informational interface between users and the privacy practice of the app and service developers~\cite{schaub_2017_des}.

Privacy policy, however, is far from an ideal interface for communicating privacy practices to users~\cite{cost, 9484434, Proctor2008}. Through a narrative literature review, we found that previous research has established that: (1) privacy policies are often lengthy~\cite{10.1145/3411764.3445465, 10.1145/1572532.1572538, 9484434, 9647767, 7020612}, ambiguous~\cite{9624976, 10.1145/3529320.3529327, 6759001, 9730898}, filled with jargon~\cite{10.1145/3411764.3445465, 6759001, 7588905}, and constantly evolving~\cite{9647767, 10.1145/3040489.3040491}; and (2) users' perceptions of privacy policy are largely negative, which precludes users from being motivated enough to understand the privacy practices and protect their privacy~\cite{9624976, 10.1145/1572532.1572538}. While these two types of challenges are closely related, existing design solutions mostly focus on the first type of challenge (e.g.,~\cite{10.1145/3529320.3529327}), leaving the critical aspect of user motivation largely unaddressed.

Undoubtedly, it is important to redesign the informational interface to effectively communicate the privacy practice of apps and services with users, particularly when such information becomes pertinent to the user's interests and rights. However, our own experience and informal discussions with other users indicate that many users still do not care about privacy or neglect it when making concrete app usage decisions, even when the information is distilled into a concise format (e.g., as privacy labels in app stores). We argue that HCI researchers need to extend their focus to motivating users to actively seek and act on related information and empower users to protect their privacy. The potential impact of redesigning the privacy policy representation can be substantially enhanced by considering this motivational issue as a primary design objective.

In this paper, we focused on addressing such a gap through a theory-driven design approach~\cite{Cash2018}. Particularly, we employed the Protection Motivation Theory (PMT)~\cite{rogers1975protection, Rogers1983} to inform design ideas specifically aimed to encourage users to take an interest in attending to their information privacy. We then used those initial design ideas to create a conceptual design prototype, enhancing the current App Store design with five features: (1) a privacy FAQ, (2) a rating on the privacy practices of the app, (3) tutorials on customizable options related to privacy, (4) a prompt for users to review privacy information, and (5) user reviews related to privacy. 

To evaluate the potential impact of our conceptual design and the underlying design ideas, we conducted an online experiment followed by a set of interviews, comparing our design with the existing App Store interface. The results showed that our design was successful in motivating users to attend to privacy. The design elements aimed to reinforce users' threat and coping appraisals (i.e., two main factors in PMT, see Section~\ref{sec:pmt}) made our participants sensitive towards their privacy in the application. In particular, participants appreciated the level of detail and organization of information in the design. They perceived such information as helpful for making better-informed decisions and taking actions regarding their personal data in the application. At the same time, to further enhance user motivation, the information needs to be condensed even more and potentially restructured. 
Overall, our research contributes to the ongoing effort to streamline privacy communication by targeting the inherent but often overlooked motivational issue related to privacy awareness. Our specific contributions are three-fold. First, we demonstrated a theory-driven design approach based on PMT that led to fundamental design concepts for motivating users to attend to privacy. Second, our approach created concrete design categories and features that can guide design efforts to directly target user motivation issues for privacy. Third, our design study and user study results provided considerations and practical implications that can inform future design.

\section{A Narrative Literature Review on the Design for Privacy Policies}
\label{sec:nlr}
We started with a narrative literature review~\cite{Baumeister1997NarrativeLR, Green2006} on the design for privacy policies to gain an extended understanding of the challenges users face when reviewing privacy policies and the existing effort on creating tools and methods to address those challenges. We focused this literature review on privacy policy because it is the fundamental, and currently most extensively investigated, informational interface between users and the companies' privacy practices. A literature review on this topic provides a feasible scope to establish practically valuable knowledge about how such an interface can be designed. Additionally, we adopted a narrative literature review approach rather than aiming to systematically synthesize all previous work published for informational interfaces for privacy; this approach enabled us to critically reflect on major related work on how to communicate privacy policy in-depth and to focus on opportunities that can be tackled by our study~\cite{Green2006}. 

\subsection{Review Process}
We conducted our search on two databases: the IEEE Xplore database and the ACM Digital Library. We chose these two databases since they both feature a wide selection of peer-reviewed works that concentrate on the design aspect of computing technology. Together, they provide high coverage of the most cited HCI venues, publishing or indexing most of the top 20 HCI journals or conferences based on h5-index~\cite{TopHCI}.

We first conducted our search using the keyword ``privacy policy'' in August 2022 to find 1061 papers in total (722 and 339 papers, in IEEE and ACM, respectively) with the publication of the papers ranging from 2007 to 2022. We then filtered the papers using their titles and abstracts following two inclusion criteria: (1) the paper must focus on privacy policies of applications or services, and (2) the paper must include the users' perspective towards privacy policies. Four authors split the task of the initial filtering; they labeled each paper as either ``include'', ``exclude'', or ``maybe,'' and noted their reasons for each decision. The four authors then discussed and resolved any uncertainties and disagreements. This initial filtering process resulted in 19 papers to include for detailed review. 

Next, we read these 19 papers, focusing on extracting the information from each paper about (1) the challenges users faced when reviewing privacy policies, and (2) the solutions and/or designs proposed to address those challenges. During this review process, three papers \cite{7977304, 9793913, 7518260} were further eliminated as their focus was not relevant to our study, resulting in the final list of 16 papers. The four authors then met and conducted an affinity diagramming~\cite{HOLTZBLATT2017127} activity to group the extracted information from the 16 papers into categories to identify common themes. 

\subsection{Review Outcome}
Table~\ref{tab:paper-summary} provides the summary of the final list of papers that underwent detailed review by the authors, as well as the identified challenges and design categories. Below, we explain how each category was discussed in representative papers. The number in parenthesis following each category indicates the number of papers that touched on that category.

\begin{table*}[t]
\centering
\small
\caption{Papers included in the narrative literature review and the extracted challenge and design categories.}
\label{tab:paper-summary}
\begin{tabular}{l|llll|ll|llllll}
\toprule
 & \multicolumn{6}{c|}{\textbf{Challenges identified}} & \multicolumn{6}{c}{\textbf{Proposed design ideas}} \\
 &  \begin{sideways} Length \end{sideways} &
    \begin{sideways} Ambiguity, vagueness \end{sideways} &
    \begin{sideways} Complexity and jargon \end{sideways} &
    \begin{sideways} Evolution and repetition \end{sideways} &
    \begin{sideways} Control autonomy \end{sideways} &
    \begin{sideways} Effort vs. benefit \end{sideways} &
    \begin{sideways} Highlighting info. \end{sideways} &
    \begin{sideways} Notification \end{sideways} &
    \begin{sideways} Privacy labels \end{sideways} &
    \begin{sideways} Other visual aid \end{sideways} &
    \begin{sideways} Navigation \end{sideways} &
    \begin{sideways} Readability assessment \end{sideways} \\
\midrule
    \citet{10.1145/1572532.1572538} (2009) & \ding{53} &  & \ding{53} &  &  & \ding{53} &  &  & \ding{53} &  &  & \\
    \citet{6759001} (2014) &  & \ding{53} & \ding{53} &  &  &  &  &  &  & \ding{53} &  & \\
    \citet{7020612} (2014) & \ding{53} &  &  &  &  &  &  &  &  &  &  & \\
    \citet{10.1145/3040489.3040491} (2015) &  &  &  & \ding{53} &  &  & \ding{53} &  &  &  &  & \\
    \citet{7588905} (2016) &  &  & \ding{53} &  & \ding{53} &  &  &  &  & \ding{53} &  & \\
    \citet{10.1145/2857705.2857741} (2016) &  &  &  &  &  &  & \ding{53} &  &  &  &  & \\
    \citet{10.1145/2872427.2883035} (2016) &  &  &  &  &  &  & \ding{53} &  &  &  &  & \\
    \citet{10.1145/3196709.3196818} (2018) &  & \ding{53} &  &  & \ding{53} &  & \ding{53} &  &  &  &  & \\
    \citet{10.1145/3230665} (2018) &  &  &  &  &  &  & \ding{53} &  &  &  &  & \\
    \citet{9647767} (2021) & \ding{53} &  &  & \ding{53} &  &  &  & \ding{53} &  &  &  & \\
    \citet{9484434} (2021) & \ding{53} &  &  &  &  &  &  &  &  &  &  & \ding{53}\\
    \citet{9624976} (2021) &  & \ding{53} &  & \ding{53} & \ding{53} & \ding{53} & \ding{53} &  &  &  & \ding{53} & \\
    \citet{10.1145/3411764.3445465} (2021) & \ding{53} &  & \ding{53} &  &  &  &  &  & \ding{53} &  &  & \\
    \citet{9606158} (2021) &  &  &  &  &  &  & \ding{53} &  &  &  &  & \\
    \citet{10.1145/3529320.3529327} (2022) &  & \ding{53} & \ding{53} &  &  &  &  &  &  & \ding{53} &  & \\
    \citet{9730898} (2022) &  & \ding{53} &  &  &  &  &  & \ding{53} &  &  &  & \\
\bottomrule
\end{tabular}
\end{table*}

\subsubsection{Challenges}
The challenges users faced when interacting with privacy policies fall into two broader groups: (1) those concerning the privacy policy itself and (2) those concerning user perception.

\addvspace{6pt}
\noindent\textbf{Challenges Concerning Privacy Policy}

\textit{Length of privacy policies ($N=5$).} 
\citet{10.1145/3411764.3445465} attributed the user experience issues of privacy policies to their length. \citet{10.1145/1572532.1572538} cited a study estimating an annual loss of productivity amounting to approximately 365 billion dollars if consumers were persuaded to read all the privacy policies they encounter. Quantifying this effort, \citet{9647767} mentioned that it would require at least 181 hours per year for a user to read the privacy policies of all used services. 

\textit{Ambiguity, vagueness, and lack of transparency ($N=5$).} 
\citet{10.1145/3529320.3529327} acknowledged that the information regarding data practices is concealed within lengthy, unclear, and ambiguous policies. \citet{9730898} discussed how services seek consent from users through the click of an ``Agree'' button but later neglect to offer the user additional information on personal data usage or sharing. \citet{9624976} described how participants exhibit distrust towards companies handling their data because of a lack of transparency in the policies regarding the process and evasive practices. 

\textit{Complexity and legal jargon ($N=5$).}
\citet{10.1145/3411764.3445465} found that privacy policies often appear too complex due to their technical language, leading users to conclude that they are not worth the reading effort. \citet{10.1145/1572532.1572538} also discussed that the readability level of typical privacy policies is equivalent to a college education. They noted that company lawyers, rather than customers, often test these policies, explaining the inclusion of complex legal language.

\textit{Evolution of policy and repetitive nature ($N=3$).}
\citet{9647767} highlighted that it is difficult for users to keep abreast with frequent changes and revisions to privacy policies. \citet{10.1145/3040489.3040491} also pointed out that all privacy policies seem identical, and mindlessly reading through them cannot contribute to understanding them.

\addvspace{6pt}
\noindent\textbf{Challenges Concerning Users' Perceptions}

\textit{Control autonomy ($N=3$).}
Users often felt a sense of powerlessness when interacting with privacy policies and protecting their data privacy. In their study, \citet{9624976} discovered that the participants believed the opt-out options to be illusory and designed in a way that does not really allow users to opt out of the service completely. Users in the exploratory design study conducted by \citet{10.1145/3196709.3196818} expressed similar frustration over how unreachable or hidden the opt-out options were.

\textit{Reasoning effort and benefit ($N=2$).} Users sometimes had difficulties assessing the benefit of attending to privacy to warrant the usually significant effort to protect their privacy. \citet{9624976} found that users are often unmotivated to read privacy policies because they do not want to devote the effort and/or just want to receive the benefits of using the service. \citet{10.1145/1572532.1572538} also pointed out that users often found it challenging to relate contents presented in privacy policies to their own use of the service.

\subsubsection{Design}
\label{subsubsec:design}
The literature reviewed helped us identify several proposed design enhancements for the usability of privacy policies. These ideas were either implemented and evaluated by the authors of the reviewed papers or suggested in the papers.

\textit{Highlighting the most relevant information ($N=7$).}
In the study of \citet{9624976}, participants suggested that important information in privacy policies should be represented in bullet points and have a larger font to make it easier to understand. \citet{10.1145/3196709.3196818} created printed booklets of privacy policies using graphic design concepts like pull quotes to highlight important points in the policy and observed that they were effective in engaging the users. Wilson et al.~\cite{10.1145/3230665, 10.1145/2872427.2883035} developed a tool to help crowd workers annotate privacy policies online, using a machine learning and natural language processing technique to highlight the most relevant paragraphs in the privacy policy for each annotation question. 
A similar idea has been explored by \citet{10.1145/2857705.2857741}. 
\citet{10.1145/3040489.3040491} developed a Javascript-based content analysis tool in which the most frequent words in the policy text are identified.

\textit{Notification of updates to policy ($N=2$).}
\citet{9647767} developed a tool that uses natural language processing to highlight key categories of sentences and changes between two versions of a policy. \citet{9730898} proposed ``Consent Receipts'' to provide evidence of user consent given for personal data usage. Their tool could alert users of changes to the policy, requiring them to re-consent and regenerate a receipt to acknowledge the changes. 

\textit{Privacy labels ($N=2$).}
The study of \citet{10.1145/1572532.1572538} represents one of the first attempts at a privacy ``nutrition'' label, through their design of Simplified Grid. They employed a combination of symbols and colors to represent privacy-related information. 
\citet{10.1145/3411764.3445465} explored different privacy policy representation formats aimed at enhancing their usability in a user study, including nutrition labels. Based on the results from the user study, the authors then developed the Visual Interactive Privacy Policy (VIPP) that added interactive elements to the nutrition labels including help icons, expandable rows, and clickable cells for additional information.

\textit{Other forms of visual aid ($N=3$).}
\citet{10.1145/3529320.3529327} created a tool called Online Privacy Policy eXplainer (PriX) that provides visual interpretations of privacy policies using privacy-related icons. \citet{7588905} conducted a usability study on the Privacy Policy Visualization Model (PPVM), designed to enhance the representation of privacy policy statements by using visual elements, symbols, and context. \citet{6759001} included visualizations in privacy policies and investigated their effects on users' trust towards internet service providers. 

\textit{Navigation ($N=1$).}
\citet{9624976} suggested embedding privacy policies directly into mobile apps to improve navigation and the overall user experience; the participants in their study found it convenient to have a sidebar delineating the various sections of the privacy policy along with a summary of the company's policy in each section. 

\textit{Readability assessment ($N=1$).}
\citet{9484434} developed a 6-point language-independent readability metric, called the Global Readability Scale, using popular metrics for English and German texts. They applied this metric to compare the readability of privacy policies from 325 mobile apps in the pre and post-GDPR periods.

\subsection{Discussion and Reflection}
Through this narrative literature review, we identified six challenges of users when interacting with privacy policies and attending to their information privacy. Four of those challenges were related to the presentation of the privacy policies themselves, while two were associated with users' perceptions of protecting their information privacy in apps and services as a whole. We also identified six design aspects explored in the literature that aimed to support users in interacting with privacy policies. Notably, these design solutions almost exclusively focused on the immediate problem of making the privacy policies easy to understand for users, mostly by improving presentation and providing shortcuts. However, the vital issue related to users' perceptions of the privacy policy, as well as their overall information privacy, is not systematically or effectively addressed in the previous work.

Besides the works reviewed above, there are other similar efforts to make information related to the privacy practice of an application or service easy to access and understand, using techniques such as contextual privacy notices~\cite{10.1145/1753326.1753561, schaub_2017_des, 10.1145/3411764.3445516} or privacy visualizations~\cite{10.1145/3600096, 10.1145/3411764.3445148}. For example, \citet{10.1145/3411764.3445516} found that salient and concise in-app privacy notices are effective in raising user awareness. Similarly, there are standardization efforts made by Apple's App Store\footnote{https://developer.apple.com/app-store/app-privacy-details/} and Google's Play Store\footnote{https://support.google.com/googleplay/android-developer/answer/10787469} requiring app developers to input privacy summaries and labels describing their application's privacy practices. However, through our preliminary discussions with users of mobile apps and online services, we found that users still do not pay attention to these privacy summaries and labels. Although they agreed with its importance when asked explicitly, they do not necessarily care to take concrete actions to protect their information privacy~\cite{DBLP:journals/popets/ZhangFYCS22}. While we acknowledge that a user may not need to engage with privacy at all times~\cite{solove}, such engagement becomes important in situations where the sensitivity of personal information is elevated and the stake of compromised privacy is high. Therefore, addressing the underlying motivational issue remains essential to make the existing effort about privacy policy presentation effective, eventually empowering users to protect their information privacy. We thus aim to contribute to this aspect and explore design guidelines and solutions that focus on motivating users to attend to their privacy when using apps and services. Particularly, we adopted a theory-driven approach to orchestrating our design process, leveraging a theoretical framework called Protection Motivation Theory (PMT)~\cite{rogers1975protection, Rogers1983} that characterizes the major factors that contribute to people's attitudes toward protection behavior (attending to their privacy, in our case).

\section{The Protection Motivation Theory}
\label{sec:pmt}
Our narrative literature review revealed that potential privacy risks and threats of using an application are inefficiently communicated through privacy policies and that we need to explore new design options to motivate users to better respond to these risks and threats when making decisions on selecting the application.
Various theoretical frameworks have been adopted to model individuals' behavior when interacting with information and communication technologies, such as Protection Motivation Theory (PMT)~\cite{rogers1975protection, floyd2000meta}, theory of planned behavior~\cite{Ajzen1985} and theory of social influence processes~\cite{doi:10.1177/001872675400700202}, among others. Given our study's focus on the motivational aspect of privacy protection, PMT stands out as an ideal theoretical framework since it specifically addresses users' protective behaviors in response to threats.

\begin{table*}[t]
\centering
\small
\caption{PMT components illustrated through an analogy of quitting smoking from the previous PMT literature~\cite{Rogers1983} and when adapted to motivate users to attend to privacy.} 
\label{tab:pmt-example}
\begin{tabular}{p{0.23\textwidth} p{0.31\textwidth} p{0.38\textwidth}} 
\toprule
PMT Component & Motivation to quit smoking~\cite{Rogers1983} & Motivation to attend to privacy \\ \midrule 
Maladaptive intrinsic rewards & Bodily pleasure, satisfaction & No need to worry about potential privacy concerns \\\addlinespace
Maladaptive extrinsic rewards & Social approval & Benefit of instantly using the app features\\\addlinespace
Perceived vulnerability & Likelihood of contracting a smoking-induced medical condition & Likelihood of the user being affected by not understanding the policy \\\addlinespace
Perceived severity & Degree of bodily harm, intrapersonal threats like self-esteem, interpersonal threats like family and work relationships &  How much the user will be affected by a privacy risk resulting from not understanding the policy \\
\midrule
Response Efficacy & Smokers consider stopping smoking to be an effective way to avoid the dangers associated with smoking & Users perceive that adopting a suggested privacy setting is an effective way to avoid privacy risks associated with not understanding the policy \\\addlinespace
Self Efficacy & Smokers think they can stop smoking & Users are confident to protect their privacy by following the suggested privacy settings \\\addlinespace
Response costs & Effort required to stop smoking & Time and effort required to understand the policy
\\\bottomrule
\end{tabular}

\end{table*}

\subsection{Protection Motivation Theory}
Protection Motivation Theory (PMT) is a theoretical framework for understanding how individuals respond to threats and subsequently engage in behaviors to protect themselves~\cite{rogers1975protection, floyd2000meta}. According to PMT, individuals engage in a cognitive process of assessing the threat and their ability to cope with it. This process involves two main factors: threat appraisal and coping appraisal. 

\textbf{Threat Appraisal} involves an individual's understanding of the amount of risk posed by engaging in a potentially harmful behavior (i.e., maladaptive response). It is influenced by the following components:

\begin{enumerate}
    \item Maladaptive Intrinsic rewards -- for potentially harmful behavior, the positive feelings or satisfaction an individual experiences when they engage in that behavior;
    \item Maladaptive Extrinsic rewards -- external reinforcements, provided by other individuals or the environment, that can encourage an individual to engage in a potentially harmful behavior;
    \item Perceived severity -- the possible degree of harm resulting from engaging in the potentially harmful behavior;
    \item Perceived vulnerability -- the probability that the individual will experience harm resulting from engaging in the potentially harmful behavior.
\end{enumerate}

\textbf{Coping Appraisal}, on the other hand, concerns an individual's assessment of their ability to change the potentially harmful behavior (i.e., adaptive response) and the ability of such a response to alleviate the threat. This factor depends on three components:

\begin{enumerate}
    \item Response efficacy -- an individual's belief that the adaptive response will truly help avoid the threat;

    \item Self-efficacy -- an individual's belief that they are able to carry out the adaptive response to avoid the threat;

    \item Response cost -- an individual's belief about how costly (e.g., in terms of money, time, effort) carrying out the adaptive response will be to the individual.
\end{enumerate}

Table \ref{tab:pmt-example} illustrates PMT through two examples, one on quitting smoking from the previous PMT literature~\cite{Rogers1983} and the other on attending to privacy in the context of our study. Among the PMT components, an increase in perceived severity and vulnerability, response efficacy, and self-efficacy, and a decrease in maladaptive rewards and response costs, can motivate behaviors preventing the threats~\cite{floyd2000meta}. Overall, PMT suggests that individuals are more likely to engage in protective behaviors when they perceive the threat as severe and likely to occur, and when they perceive the available coping strategies as effective and feasible.

\subsection{Applications of PMT in HCI, Security and Privacy}

Originally applied to the healthcare domain, PMT and its variations have found numerous applications in fields related to computing, including the study of protective behaviors related to online privacy threats~\cite{doi:10.1080/02650487.2016.1239878, 10.1145/3491102.3517643, posey} and information-sharing and security behaviors~\cite{Sommestad2014AMO, doi:10.1080/15252019.2018.1521317, 10.5555/3235924.3235929, 10.5555/3291228.3291232, anderson, 10.1016/j.cose.2011.10.007, 6480204, doi:10.1080/02681102.2013.814040}. PMT has also been applied to develop nudging interventions to encourage users to adopt secure behaviors~\cite{255664, 10.2478}. Moreover, \citet{Chennamaneni2022} used PMT to develop a research model to investigate the factors influencing mobile app users' privacy concerns and their corresponding privacy protection behaviors. \citet{Ganesh2022} used PMT as a base to build a 2D persuasive game that instructed users about privacy and security best practices in smartphones. In our work, we consider PMT as the framework to structure the design ideation and we further used it to assess the impact of the design ideas on users' motivation in a user study.
\section{Design Process}
Following a theory-driven iterative process~\cite{Cash2018}, we proposed a conceptual design for the product page of mobile apps in app stores, in particular, for a hypothetical fitness application as a concrete case study. We first generated design ideas targeting different components in PMT that might impact human motivation when preventing negative outcomes of certain behaviors. We then iteratively created preliminary conceptual designs and refined them by incorporating user feedback from initial user studies.

\subsection{Case Study Selection}
Our study focuses on redesigning the product page for applications in mobile app stores. We selected the iOS Apple Store as our target given the popularity of iOS as a mobile operating system, but the design ideas presented are applicable to other app stores as well, such as Google Play Store. The product page serves as the initial user touchpoint, providing primary information about the application, such as user ratings, key features, categories, etc. Currently, the privacy practices of an application are disclosed under the App Privacy section in the format of privacy labels. These labels are grouped based on whether the data may be used to track an individual across other apps and websites or whether the data is linked to one's identity. We consider the current design of the App Store product page as a baseline and enhance it with more privacy-related information for the application corresponding to the different design categories.

For the conceptual design prototype, we decided to create a mock application in the category of fitness apps. Fitness apps, that encourage users to be physically active, are popular and widely downloaded~\cite{YIN2022107106}. However, those apps collect and deal with a large amount of sensitive information, such as biometric data (e.g., heart rate, step count), location data, and personal health information (e.g., weight, diet, sleep patterns). While such information can provide detailed insights into a user's daily routine, health status, and personal habits, the sensitive nature of those data makes fitness apps particularly attractive to hackers and data brokers seeking to exploit it for nefarious purposes~\cite{Velykoivanenko2022, Dhondt2022}.

We examined ten popular fitness apps on the App Store (Adidas Running, Apple Health, Asana Rebel, Google Fit, Fitbit, Komoot, FITIV Pulse, Nike Training Club, Peloton, and Strava) to understand their privacy-sensitive processes. Among them, Strava and Nike Training Club collected the widest variety of personal information, including identifiers, purchase data, usage data, location, health and fitness data, contact information, financial information, and search history, among others. By studying the privacy policies of the two applications in greater detail, as they represent the most severe privacy concerns, we derive the privacy practices for our mock application, named UFit.

\subsection{Design Ideation}
\label{subsubsec:ideation_proc}
The ideation process started with brainstorming sessions, during which all authors of this work generated and discussed design ideas targeting each component from PMT. We then clustered the ideas into six high-level categories based on the design tactics and further generated multiple design ideas in each category. Table~\ref{tab:shortlist_designideas} summarizes these categories and the design ideas that emerged from brainstorming.

\begin{table*}[t!]
\centering
\small
 \caption{Design categories and ideas emerged from the design ideation phase} 
\label{tab:shortlist_designideas}
% \resizebox{\linewidth}{!}{
\begin{tabular}{p{0.1\linewidth} p{0.25\linewidth} p{0.20\linewidth} p{0.35\linewidth}} 
\toprule
\textbf{Category} & \textbf{Description} & \textbf{Associated PMT Comp.} & \textbf{Design Ideas*} \\ \midrule
Assistance & Providing assistance and scaffolding to users in managing their information privacy can enhance their confidence and sense of control. &
    \begin{minipage}[t]{\linewidth}
        \begin{itemize}[leftmargin=*]
            \item Self-Efficacy
            \item Response Efficacy
        \end{itemize}
    \end{minipage}&
    \begin{minipage}[t]{\linewidth}
        \begin{itemize}[leftmargin=*]
            \item[{\faObjectUngroup[regular]}] Introducing ``privacy tutorial'' to help the user understand how to attend to their information privacy in the application.
            \item[{\faLightbulb[regular]}] Providing flexible options and making it convenient for users to adjust and customize their privacy settings. %\cite{10.1145/3411764.3445465}
            \item[{\faLightbulb[regular]}] Highlighting the available choices and options within the privacy policy to improve user awareness.
        \end{itemize}
    \end{minipage}\vspace{2pt}\\ \midrule
    
Condensing Information & Summarizing the privacy policy and privacy-related information into more relevant and digestible textual formats to reduce information overload &
    \begin{minipage}[t]{\linewidth}
        \begin{itemize}[leftmargin=*]
            \item Response Cost
            \item Response Efficacy
        \end{itemize}
    \end{minipage}&
    \begin{minipage}[t]{\linewidth}
        \begin{itemize}[leftmargin=*]
            \item[{\faObjectUngroup[regular]}] Using FAQs to summarize the privacy policy and discuss other common privacy-related topics.
            \item[{\faLightbulb[regular]}] Summarizing the privacy policy using plain language and explanations. %\cite{10.1145/2857705.2857741, 217470, 10.1145/3491102.3517688}
        \end{itemize}
    \end{minipage}\vspace{2pt}\\ \midrule

Data \& Statistics & Presenting privacy-related data and statistics to offer a more concise representation of privacy-related information. &
    \begin{minipage}[t]{\linewidth}
        \begin{itemize}[leftmargin=*]
            \item Perceived Vulnerability
            \item Perceived Severity
        \end{itemize}
    \end{minipage}&
    \begin{minipage}[t]{\linewidth}
        \begin{itemize}[leftmargin=*]
            \item[{\faObjectUngroup[regular]}] Including ``privacy rating'' to convey the level of privacy-related risks associated with downloading the app. %\cite{Tsai2011}
            \item[{\faObjectUngroup[regular]}] Presenting privacy violation cases for similar apps in the category.
            \item[{\faLightbulb[regular]}] Indicating the number of users who viewed or used the opt-out options.
        \end{itemize}
    \end{minipage}\vspace{2pt}\\ \midrule

Inhibition & Encouraging or mandating the users to attend to privacy before allowing them to use the app features. &     
    \begin{minipage}[t]{\linewidth}
        \begin{itemize}[leftmargin=*]
            \item Maladaptive Rewards
        \end{itemize}
    \end{minipage}&
    \begin{minipage}[t]{\linewidth}
        \begin{itemize}[leftmargin=*]
            \item[{\faObjectUngroup[regular]}] Prompting the user to first read the privacy policy when clicking the Get button to download the app.
            \item[{\faLightbulb[regular]}] Hiding positive reviews until the privacy policy is read.
            \item[{\faLightbulb[regular]}] Reducing the value of the service until the privacy policy is read.
            \item[{\faLightbulb[regular]}] Requiring a minimum time spent on the privacy policy before download.
        \end{itemize}
    \end{minipage}\vspace{2pt}\\ \midrule

Social Interaction & Making users aware of privacy problems in the application through sharing and collaboration. &
    \begin{minipage}[t]{\linewidth}
        \begin{itemize}[leftmargin=*]
            \item Perceived Vulnerability
            \item Perceived Severity
            \item Self-Efficacy
        \end{itemize}
    \end{minipage}&
    \begin{minipage}[t]{\linewidth}
        \begin{itemize}[leftmargin=*]
            \item[{\faObjectUngroup[regular]}] Showing social signals such as privacy-specific user reviews to emphasize privacy concerns raised by other users.
            \item[{\faLightbulb[regular]}] Allowing users to contribute to the understanding of privacy policies by providing crowd annotations and comments. %\cite{10.1145/2872427.2883035, 10.1145/3230665}
        \end{itemize}
    \end{minipage}\vspace{2pt}\\ \midrule
    
Visualization & Simple and intuitive visualizations can convey complex privacy concepts in a more accessible and engaging manner. &
    \begin{minipage}[t]{\linewidth}
        \begin{itemize}[leftmargin=*]
            \item Perceived Vulnerability
            \item Perceived Severity
            \item Response Cost
        \end{itemize}
    \end{minipage}&
    \begin{minipage}[t]{\linewidth}
        \begin{itemize}[leftmargin=*]
            \item[{\faObjectUngroup[regular]}] Using icons to reduce information overload and aid understanding. %\cite{10.1145/3411764.3445387}
            \item[{\faLightbulb[regular]}] Highlighting the risks and threats in the privacy policy and user reviews %\cite{10.1145/3196709.3196818}.
            \item[{\faObjectUngroup[regular]}] Using visual cues (color, size, etc.) to indicate the severity. %\cite{9624976}
            \item[{\faLightbulb[regular]}] Using comics or fiction to encourage imagining potential privacy risks. %\cite{10.1145/3411764.3445465}
        \end{itemize}
    \end{minipage}\vspace{2pt}\\

\bottomrule
\end{tabular}
% }
{\raggedright $^{*}$ \small
 {\faObjectUngroup[regular]} \textit{Ideas emerged from brainstorming and included in our conceptual design} \newline
\hspace*{0.4mm} {\faLightbulb[regular]} \textit{Ideas emerged from brainstorming but not included in the conceptual design due to feasibility or complexity}
 \par}
\end{table*}

\subsection{Design Iterations}
An iterative approach was followed for the conceptual design of the App Store product page of our mock application UFit. We first developed our initial design ideas presented in Table \ref{tab:shortlist_designideas} into features using Figma\footnote{https://www.figma.com/} for critique and iteration among the authors. We initially created separate features to capture each design idea. We then conducted a user study with three participants to collect early user feedback and further iterate the design. The participants were recruited from our personal network and compensated \$30 CAD. All participants were female and current university students.
During the study, we asked their opinions, as potential users of UFit, on the perceived usefulness of each privacy-related feature we proposed, how the features motivate them to understand and protect their privacy, and their other feedback related to the design. 

The primary feedback was regarding merging and complementing closely related features, increasing the prominence of features, and avoiding overloading information. We used these insights to make further changes to the design of the conceptual prototype, notably combining the originally separated features into a coherent design. Particularly, we moved the privacy violation cases under the privacy rating page (both related to the \textit{Data \& Statistics} category, see Table~\ref{tab:shortlist_designideas}) since they are closely related. We also added two prominent buttons on top of the app screenshots to make easy access to the Privacy FAQ (from the \textit{Condensing Information} category) and the Privacy Tutorial (from the \textit{Assistance} category). In addition, we adjusted the feature related to \textit{Inhibition} to direct the user to the Privacy FAQ instead of the original, hard-to-digest privacy policy. Similarly, the design ideas for the categories \textit{Visualization} and \textit{Condensing Information} were combined to link the icons with the information in the FAQ instead of the privacy policy. In the next section, we elaborate on the details of our final design.

\section{Final Design}
\begin{figure*}
    \centering
    \includegraphics[width=\linewidth]{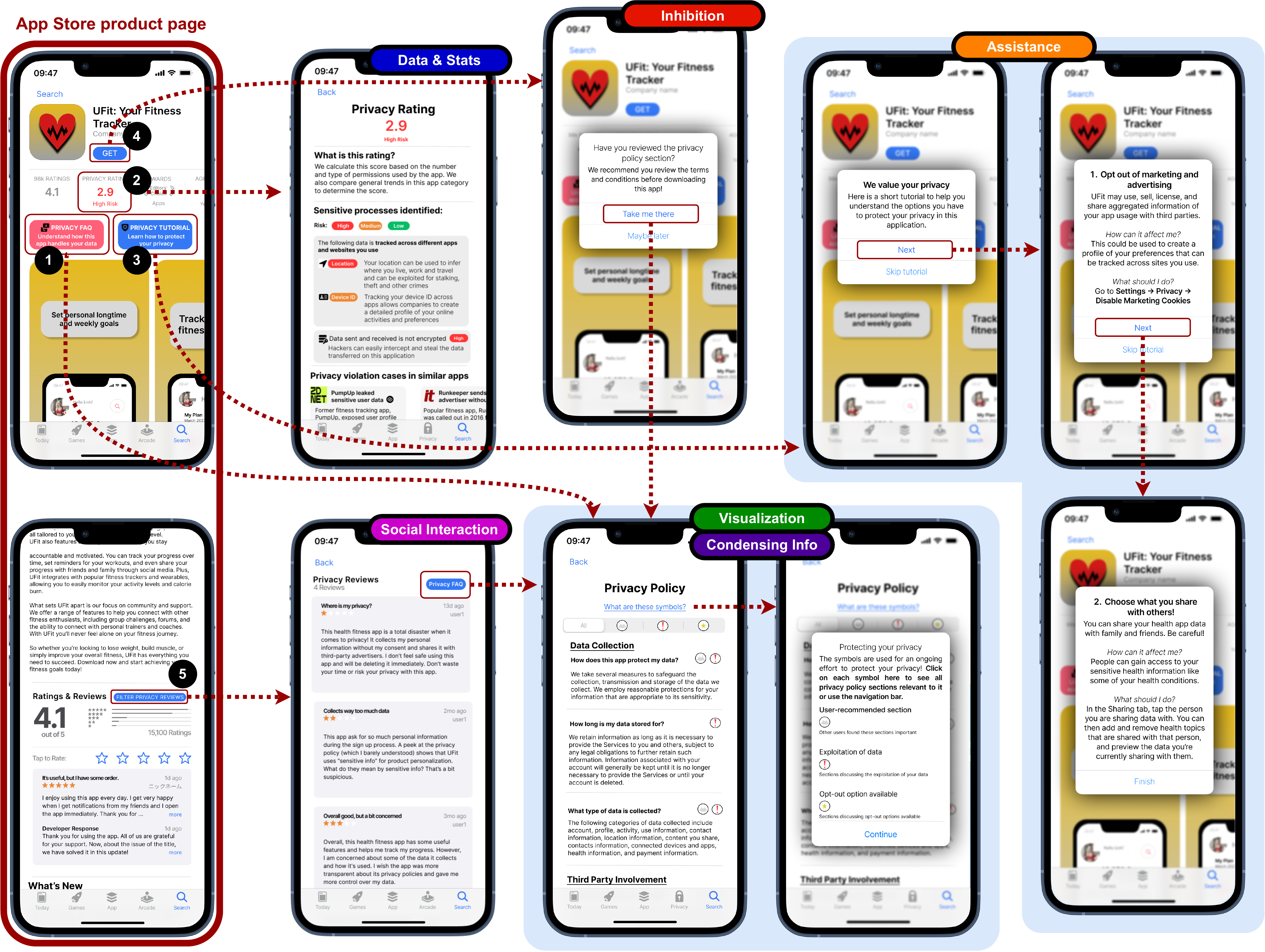}
    \caption{Primary features to motivate privacy consideration in our final design, including \circled{1} privacy FAQ, \circled{2} rating on privacy practice of the app, \circled{3} a short tutorial on customizable privacy options, \circled{4} a prompt for users to review privacy FAQ before downloading the app, and \circled{5} user reviews related to privacy.}
    \label{fig:overall_design}
    \Description{An array of 10 Figma design iPhone screens showcasing various user interface designs and interactions built around our six design categories: Data and Statistics, Inhibition, Assistance, Social Interaction, Visualization, and Condensing Information. Shown on the main App Store homepage for the app, there are 5 main design implementations explored: 1. a pink button leading to the privacy FAQ, 2. a red color rating section with a number indicating the privacy rating that leads to the privacy rating section, 3. a blue button that leads to the privacy tutorial on customizable privacy options, 4. The standard ``Get'' button on the App Store leads to a pop-up prompt for users to review the privacy FAQ before downloading, 5. A blue button that leads to additional user reviews related to privacy.}
\end{figure*}

The final design of the product page in the App Store incorporated five main features across the six design categories in Table~\ref{tab:shortlist_designideas}: (1) \textit{privacy FAQ}, (2) \textit{rating on privacy practice}, (3) \textit{tutorials on customizable options}, (4) \textit{a prompt for users to review privacy FAQ}, and (5) \textit{user reviews related to privacy}.  Figure~\ref{fig:overall_design} illustrates how users can interact with these features when encountering an application.

\subsection{Privacy FAQ} 
A prominent red button at the top of the product page directs the user to Privacy FAQ (\circled{1} in Figure~\ref{fig:overall_design}). This section presents information from the privacy policy~\cite{zhang23} and additional information in a compact question-answer format, divided into three main sections, i.e., \textit{Data Collection} (e.g., what type of data is collected and how long is my data stored for), \textit{Third Party Involvement} (e.g., what third parties are involved and can I control what data is shared with third parties), and \textit{User Options} (e.g., what opt-out options are available and can I adjust my settings anytime). 

Additionally, each question in the FAQ is annotated with icons representing the following properties of the information: (1) User recommended, indicating a scenario wherein users mark certain sections as important or must-read; (2) Exploitation of data, indicating potential misuse of personal data; and (3) Opt-out options, discussing the options available for a user to opt out of having their personal information collected, shared or used in a certain way. Users can filter information by those properties easily by clicking the corresponding icons on top of the FAQ. 

\subsection{Privacy Rating}
\label{subsec:privacy_rating}
Privacy rating mimics the concept of app ratings present in the App Store, serving as an assessment report on the app's level of adherence to privacy best practices and data protection standards. In our design, the privacy rating is presented next to the app rating in the top bar of the product page. The text indicating the risk level (i.e., high, medium, low) is colored-coded (red, yellow, or green respectively) to draw the user's attention. Clicking on the rating takes the user to a page giving more information about the rating (\circled{2} in Figure \ref{fig:overall_design}). This page gives a brief description of the privacy rating and lists the sensitive processes (individually ranked by risk level) contributing to the rating, such as the permissions accessed by the app, information collected and shared with third parties, or questionable privacy practices in the application.

At the bottom of the privacy rating page is a section pertaining to cases of privacy violations in similar applications. Each case is presented in a card containing a title, a brief summary of the incident, and a link to a relevant article or external source. This section aims to help the user get familiar with how their information privacy could be compromised through relevant real-world incidents and therefore contextualize the presented privacy rating.

\subsection{Privacy Tutorial}
The features related to configuring the privacy options in an application are often hard to locate and, even when located, difficult for the users to understand the consequences~\cite{10.1145/3411764.3445465}. The privacy tutorial in our design presents users with guidance on how to take actions related to their information privacy in the app. Our tutorial highlights the primary concerns related to the privacy practices of the app, inspired by the privacy violation cases and common fitness app features. The first part of the tutorial describes how the privacy practice of the app would impact the users; then, it suggests step-by-step instructions on locating and setting the appropriate privacy options to mitigate the risk (see \circled{3} in Figure~\ref{fig:overall_design}).

\subsection{Privacy Prompt}
Most mobile applications today require users to agree to the terms of use and privacy policy after downloading the application, generally during the registration flow. However, it can be beneficial to nudge users towards examining the privacy information of the application before downloading it to help them make more privacy-aware decisions~\cite{255664, 10.2478}. Our design incorporated a pop-up that is displayed when the user clicks the ``Get'' button to download the application, encouraging the user to visit the Privacy FAQ page before they download the application (\circled{4} in Figure~\ref{fig:overall_design}).

\subsection{Privacy Reviews}
User ratings and reviews are among the most important factors that shape the decision-making process of online consumers~\cite{cheung2012impact}. While app stores have already integrated user ratings and reviews, the sheer number of reviews left for popular apps makes it challenging to draw users' attention to reviews concerning privacy. In our design, we included a button under this section to extract the privacy-related reviews (\circled{5} in Figure \ref{fig:overall_design}). Clicking on the button takes the user to a page displaying only those reviews sorted in reverse chronological order. A review is classified as privacy-related if it discusses topics such as the permissions used by the application, personal data collected, used, or shared with third parties, and the privacy practices of the application.

\section{User Study Methods}
We conducted a user study to understand the potential impacts of our conceptual design and collect user feedback on specific design ideas. The user study included an online experiment and a follow-up interview. The Research Ethics Board of all involved institutions approved the study. Below, we outline the data collection and analysis methods for each part of our study.

\subsection{Online Experiment}
We conducted a between-group online experiment to assess the effects of two designs on the privacy awareness and motivation of the participants: (1) our conceptual design of the enhanced App Store product page as shown in Fig.~\ref{fig:overall_design} (treatment condition) and (2) a mockup of the current iOS App Store design containing the App Privacy section (control condition).

\subsubsection{Participants}
Participants were recruited from Prolific\footnote{https://www.prolific.co}. Participants had to meet the following criteria to access our study: (1) they can conduct the study in English, (2) they use mobile devices at least 2 to 3 times per day, and (3) they have completed 10 previous Prolific submissions with a minimum acceptance rate of 85\%. Each participant was compensated with \$5 CAD.

In total, 87 participants completed the online experiment in the treatment condition and 85 in the control condition. Among the completed submissions, 11 were rejected due to low-quality input (six from the treatment group, and five from the control group). The quality of the input was deemed low if: (1) the participant did not complete a required section of the experiment (such as completing the exploration of the design) or (2) the participant demonstrated low-effort responses (such as one-word answers to all open-ended questions).

Among the remaining 81 treatment group participants and 80 control group participants, around 50\% were between 18 and 24 years of age, and approximately 55\% were male. The highest proportion of participants had a bachelor's degree in both conditions (51.9\% for the treatment condition and 35\% for the control condition). A total of 33 participants indicated that they used mobile devices every day, while 128 used mobile devices multiple times every day. The detailed demographics are displayed in Table~\ref{tab:online-participants}.

\begin{table*}[th]
\centering
\small
\caption{Participant demographics of the online experiment} 
\label{tab:online-participants}
\begin{tabular}{ p{0.10\linewidth}  p{0.30\linewidth} p{0.1\linewidth} p{0.1\linewidth} } 
\toprule
& & Treatment & Control \\ \midrule
Age group & 18-24 years \hfill \break 25-34 years \hfill \break 35-44 years \hfill \break 45-54 years \hfill \break 55-64 years \hfill \break 65+ years & 36 \hfill \break 31 \hfill \break 7 \hfill \break 2 \hfill \break  4 \hfill \break 1 & 41 \hfill \break 23 \hfill \break 12 \hfill \break 4 \hfill \break  - \hfill \break - \\ \vspace{0.2mm}
Gender & \vspace{0.2mm} Male \hfill \break Female \hfill \break Non-binary / third gender & \vspace{0.5mm} 44 \hfill \break 36 \hfill \break 1 & \vspace{0.2mm} 48 \hfill \break 31 \hfill \break 1 \\ \vspace{0.5mm}
Education & \vspace{0.5mm} Less than high school degree \hfill \break High school graduate \hfill \break Some college / associate degree \hfill \break Bachelor's degree \hfill \break Master's degree \hfill \break Doctoral degree & \vspace{0.5mm} - \hfill \break 10 \hfill \break 19 \hfill \break 42  \hfill \break 9 \hfill \break 1 & \vspace{0.5mm} 2 \hfill \break 25 \hfill \break 18 \hfill \break  28 \hfill \break 6 \hfill \break 1\\
\bottomrule
\end{tabular}
\end{table*}

\subsubsection{Experiment procedure}
\label{sec:online_study_procedure}
The procedure and material of the experiment (e.g., questionnaire) were refined through multiple rounds of in-person and online pilot studies. During the actual experiment, after answering demographic questions, participants were asked to consider the scenario of deciding on a fitness app to install and were directed to explore a UFit product page design hosted on Maze\footnote{https://maze.co}. According to the participant's experimental group, either our conceptual design (see Fig.~\ref{fig:overall_design}) or a mockup of the current App Store design was shown to the participant. When the participants indicated that they completed exploring the product page, they were prompted with a message asking them if they had visited all privacy-related sections; the treatment condition prompted a message that links to the four main design concepts (i.e., privacy FAQ, privacy rating, privacy tutorial, and privacy reviews), while in the control condition, this message included a link to the App Privacy section of the page. %We acknowledge that this is slightly different from the natural setting. 
The prompt was added based on the feedback from the pilot studies to ensure that the participants have the chance to be exposed to privacy-related information \textit{in both conditions}. This study design aligns with our objective to understand whether the proposed design can effectively motivate users, compared with the current App Store design of privacy labels. For the same reason, participants were made aware of the study objective through the study title when giving their consent to join the study. 

After exploring the product page design, participants in both conditions were asked to rate their level of agreement on a 5-point Likert scale for five statements, listed in Table~\ref{tab:PMT-statements}, each corresponding to one PMT dimension. For each rating, they were also asked to provide a textual explanation of their rating. %Participants were not introduced to the concept of the PMT dimensions. 
The statements were inspired by the work of \citet{Chennamaneni2022} and refined through the pilot studies to ensure their clarity; the statement for maladaptive rewards was omitted as the users in our pilot study indicated confusion about the underlying concept. To minimize potential order effects in the study, we implemented a counterbalancing strategy by reversing the sequence of the PMT questions for half the participants in both conditions. After completing an equal number of control and treatment studies with the reversed question order, we did not find a significant difference in the ratings with different question orders. Our data analysis is thus based on the merged samples of questions in original and reversed orders.

Finally, before exiting the experiment, participants were asked to provide overall feedback on the design. They were also asked to state their interest in a follow-up interview. Among the 161 accepted submissions, the median completion time of the experiment was 15.34 minutes.

\begin{table*}[th]
\centering
\small
\caption{Final PMT statements for online experiment} 
\label{tab:PMT-statements}
\begin{tabular}{p{0.73\linewidth}  p{0.21\linewidth} } 
\toprule
Statement & Corresponding PMT Comp. \\ \midrule
- This design helps me understand the likelihood of privacy-related problems that I could face when using this application. & Perceived Vulnerability \\ 
- This design helps me understand the severity of privacy-related problems that I could face when using this application. & Perceived Severity \\ 
- The features presented in the design can help me attend to the privacy-related problems in this application. & Response Efficacy \\ 
- I am confident that I can take actions to attend to my information privacy in the application through this design. & Self Efficacy \\ 
- This design reduces the effort required to attend to my information privacy in this application. & Response Cost\\ \bottomrule
\end{tabular}
\end{table*}

\subsubsection{Data analysis}
We conducted a Mann-Whitney U test on each PMT dimension to assess if there was a difference between the treatment group and the control group. Then, for the explanations that the participants provided in both conditions, we conducted an inductive thematic analysis~\cite{Vaismoradi2013} using Atlas.ti to identify themes in participants' positive and negative responses related to the PMT dimensions; the codes and themes were organized by the type of design (i.e., control or treatment). Particularly, two researchers first independently coded the responses from the treatment and control groups. The coding was then discussed among all four authors until any disagreements or concerns were resolved.

\subsection{Interview}
We invited participants who had indicated further interest during the online experiment to participate in a follow-up interview.

\subsubsection{Participants}
We prioritized recruiting participants who provided high-quality responses during the online experiment -- they had descriptive answers in the survey and shared opinions with the prospect of further elaboration. This recruitment method was used to increase the likelihood of gathering more constructive insights. A total of nine participants participated in the interviews: five from the control condition and four from the treatment condition. Four of the interview participants were female and five were male. Their age ranged from 18 to 34. Table \ref{tab:interview-participants} summarizes the participants.

\begin{table*}[t]
\centering
\small
\caption{Participant demographics of the interview} 
\label{tab:interview-participants}
\begin{tabular}{llllll} 
\toprule
ID & Gender & Age Group & Education & Design tested in online experiment & Weekly device usage\\ \midrule 
P1 & Male & 18-24 & Bachelor's degree & Control & Every day\\
P2 & Male & 18-24 & Bachelor's degree & Treatment & Multiple times every day\\
P3 & Female & 25-34 & Bachelor's degree & Control & Every day\\
P4 & Male & 25-34 & Bachelor's degree & Control & Multiple times every day\\
P5 & Male & 18-24 & Bachelor's degree & Control & Multiple times every day\\
P6 & Female & 18-24 & Some college/ associate degree & Treatment & Multiple times every day\\
P7& Male & 25-34 & High school graduate & Control & Every day\\
P8 & Female & 25-34 & Some college/ associate degree & Treatment & Multiple times every day\\
P9 &Female & 18-24 & Bachelor's degree & Treatment & Multiple times every day
\\\bottomrule
\end{tabular}
\end{table*}

\subsubsection{Interview procedure}
All interviews were held on Zoom and were recorded. Each interview lasted around 30 minutes and each participant was compensated with \$15 CAD. We asked them personalized questions about some of their insightful responses to the survey in the online experiment. For example, if a participant suggested an improvement in our design, we would ask them to elaborate on their ideas and rationales. Following that, we asked participants to explore the design prototype they had not seen during the experiment (i.e., participants from the control condition would see the treatment design, and vice versa). We then asked them to rate the PMT questions for the new design and compare it to the original design they saw in the experiment. This interview study thus aimed to enrich our understanding of the rationales behind the participants' online experiment responses and allow us to better compare the potential impacts of both designs.

\subsubsection{Data analysis}
All interviews were transcribed verbatim. Then, through thematic analysis~\cite{Vaismoradi2013}, we refined and enriched the themes identified from the qualitative answers from the online experiment. The codes and themes were used to enrich those that emerged from the survey, as the interview was designed to allow the participants from the online experiment to elaborate on their qualitative responses provided in the survey.

\section{User Study Results}
Our findings indicate that the treatment design was particularly successful in increasing participants' \textit{Perceived Vulnerability} and \textit{Perceived Severity}, but fell short of reducing \textit{Response Cost}. Participants' qualitative feedback provided useful insights into the reasons behind this outcome. In the sections below, we first report the statistical results of participants' ratings on PMT dimensions and then describe the themes we identified in their feedback on both the control and treatment designs.

\label{sec:results}
\subsection{Rating on PMT Dimensions}

\begin{figure*}[t]
    \centering
    \begin{subfigure}{0.19\textwidth}
    \centering
    \includegraphics[width=\linewidth]{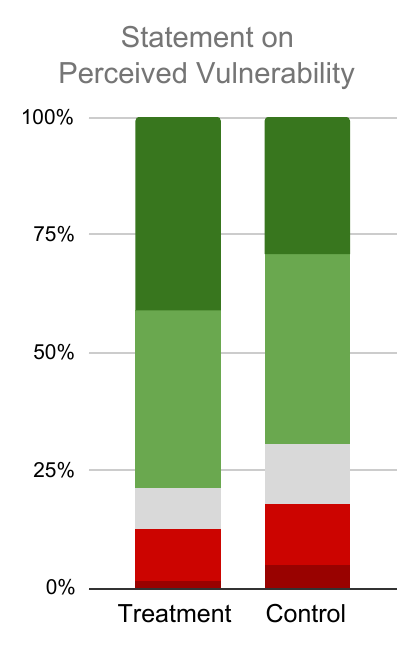}
  \end{subfigure}
  \begin{subfigure}{0.19\textwidth}
    \centering
    \includegraphics[width=\linewidth]{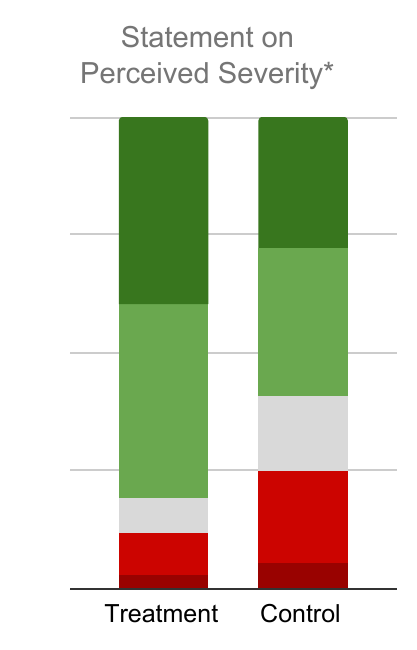}
  \end{subfigure}
  \begin{subfigure}{0.19\textwidth}
    \centering
    \includegraphics[width=\linewidth]{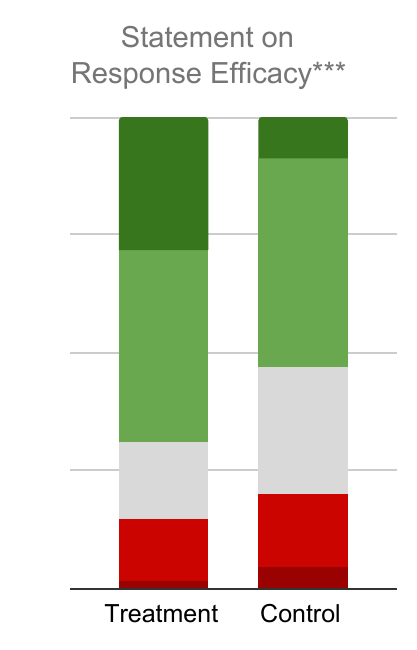}
  \end{subfigure}
  \begin{subfigure}{0.19\textwidth}
    \centering
    \includegraphics[width=\linewidth]{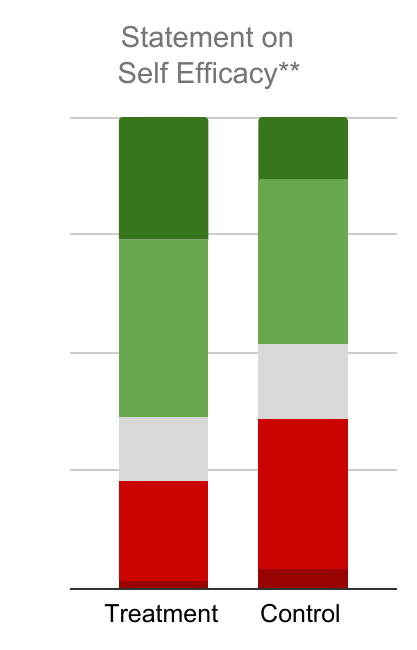}
  \end{subfigure}
  \begin{subfigure}{0.19\textwidth}
    \centering
    \includegraphics[width=\linewidth]{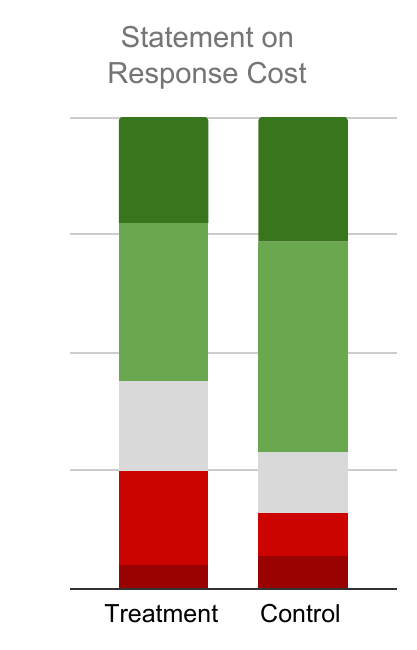}
  \end{subfigure}
    \begin{subfigure}{0.55\textwidth}
    \centering
    \includegraphics[width=\linewidth]{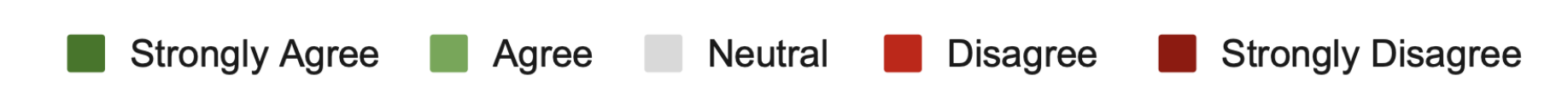}
    \end{subfigure}
    \caption{Participant ratings for PMT statements. The stars indicate significant differences between the control and the treatment conditions (*** p < 0.001, ** p < 0.01, * p<0.05).}
    \Description{Five bar graph charts displaying treatment and control results for each PMT statement in the following order: Statement on Perceived Vulnerability, Statement on Perceived Severity, Statement on Response Efficacy, Statement on Self Efficacy, Statement on Response Cost. The distribution of responses from Strongly Disagree, Disagree, Neutral, Agree, and Strongly Agree are shown for each statement based on participant input. For the treatment design, Perceived Severity received the highest percentage of positive ratings (Agree and Strongly Agree) with 80.7\% and Response Cost received the least percentage of positive ratings with 56.3\%. 78.8\% of participants agreed that the treatment increased their Perceived Vulnerability, while 69.1\% of the ratings for Response Efficacy and 63.8\% of the ratings for Self Efficacy were positive. Meanwhile, the control design received the highest percentage of positive ratings for Response Cost with 71.0\% and the least percentage of positive ratings for Self Efficacy with 48\%. 69.6\% of participants agreed that the control increased their Perceived Vulnerability, while 59.2\% of the ratings for Perceived Vulnerability and 52.9\% of the ratings for Response Efficacy were positive.}
    \label{fig:fig3}
\end{figure*}

Fig.~\ref{fig:fig3} shows the results of the PMT ratings for both the treatment and control designs. For all PMT components, except \textit{Response Cost}, the treatment design received a higher percentage of positive ratings (Agree and Strongly Agree) as compared to the control. The treatment design was most effective in addressing \textit{Perceived Vulnerability} and \textit{Perceived Severity}, as they received the highest number of positive ratings (78.8\% and 80.7\% respectively); in comparison, the control design received positive ratings of 69.6\% and 59.2\% for those same categories respectively. However, participants felt that the control design targeted \textit{Response Cost} effectively (71.0\% positive ratings) compared to the treatment design (56.3\% positive ratings).

Mann-Whitney U tests (with the two-tailed hypothesis and alpha level of 0.5) were carried out for each of the PMT components to compare the results between the treatment and control groups. The tests showed that the differences were statistically significant for \textit{Perceived Severity} ($U=2521$, $p=0.015$), \textit{Response Efficacy} ($U=2128$, $p<0.001$) and \textit{Self-Efficacy} ($U=2428$, $p=0.006$). Meanwhile, the differences between the groups for \textit{Perceived Vulnerability} ($U=2756$, $p=0.103$) and \textit{Response Cost} ($U=2694$, $p=0.352$) were not significant.

During the interview, the participants were explicitly asked which design they thought was more effective in motivating users to attend to their information privacy. Three participants felt that the control was better in this regard, while the other six participants agreed that the treatment design was more effective. Their reasons aligned with the themes that we describe below.

\subsection{User Feedback on the Control Design}
Our thematic analysis revealed various themes regarding the positive and negative aspects of the control design (i.e., the original App Store product page with the privacy labels) that affected the participants' perception of the PMT dimensions. The number $N$ after each theme refers to the number of participant comments from the online experiment relating to that theme.

\subsubsection{Positive factors about the control design} \hfill

\textbf{Concise and simple} ($N=98$):
Participants considered the privacy labels included in the App Store to be brief, compact, and straightforward. When discussing \textit{Perceived Vulnerability} and \textit{Perceived Severity}, participants mentioned that \textit{``[The design] is compact and to the point''} and \textit{``the icons are easier to look at than sentences.''} The lower amount of text in the design also increased participants' perceived \textit{Response Efficacy} and \textit{Self Efficacy}, as P6 commented during the interview: \textit{``The information is summarized and easy to read because you know where to look.''} Similarly, the perceived \textit{Response Cost} also decreased; e.g., \textit{``Everything is placed in a manner that is easy for the user to analyze those statistics.''}

\textbf{Facilitate informed decision-making} ($N=59$):
Participants noted that the privacy labels guided their decision-making process on whether to download the app. For example, a participant expressed a \textit{Perceived Vulnerability}: \textit{``This design ... makes me aware of just how much data/information the app needs to access. ... Unless I could change these settings in the app I wouldn't download it.''} Another participant also mentioned when discussing \textit{Perceived Severity}: \textit{``If an app asks for an alarming amount of information access, I can choose not to install it.''} For both \textit{Response Efficacy} and \textit{Self Efficacy}, participants felt that the control design allowed them to understand available privacy-related options, helping them decide how to use the app; for example, \textit{``There seem to be sufficient options related to data collection.''} Participants also mentioned that the control design reduced \textit{Response Cost} of making decisions; e.g., \textit{``You get a good primer on what there is and is a good option for making a preliminary decision.''}

\textbf{Broad coverage of privacy risks} ($N=52$):
Participants thought that the App Store's app privacy section offered comprehensive information about potential privacy risks associated with using the app. For example, when discussing \textit{Perceived Vulnerability}, a participant mentioned: \textit{``Every data that could get stolen/leaked etc. is mentioned in the app privacy section.''} When talking about \textit{Perceived Severity}, a participant also wrote: \textit{``The design clearly shows me what types of data are used for which purposes and might therefore lead to privacy-related problems I am uncomfortable with.''} Similarly, users felt that the thorough information allowed them to confidently make informed decisions when downloading the app, increasing their \textit{Response Efficacy} and \textit{Self Efficacy}, as a participant indicated: \textit{``being transparent about the data gives me the opportunity to make conscious choices.''}.

\subsubsection{Negative factors about the control design.} \hfill

\textbf{Overly condensed} ($N=61$):
Many participants mentioned that the information presented lacks details and clarity, as P9 articulated during the interview, \textit{``It's just telling you like `this may be collected' but not telling you how to avoid it or any implications of the sensitive information being tracked''}, decreasing \textit{Perceived Vulnerability}. Another participant expressed a lower \textit{Perceived Severity} as the design \textit{``did not explain in detail how severe these different privacy options were.''} A participant also wrote about \textit{Response Efficacy}: \textit{``the design is merely informative and is oblivious to any problems with the application.''} At the same time, participants' \textit{Self-Efficacy} was negatively affected by the overly condensed information, as a participant explained that \textit{``the next step has to be wading through the privacy agreements... to understand how that data is being gathered and to what end... And this can be a laborious task.''} To improve the design, participants suggested that there should be \textit{``clearer and more lucid descriptions''} that uncover \textit{``what may be hidden by the app developer.''} A participant also specified their information needs: \textit{``I would like to be able to get more inside info as to why this kind of data and what for it is going to be collected. How is it related to fitness and in what way does it improve the final product's user experience.''} Some participants suggested adding Q\&As and external sources to address this.

\textbf{Lack of actionable options and features} ($N=51$):
Many participants felt that the privacy section provided information but lacked options to help address their privacy concerns, leading to lower \textit{Response Efficacy} and \textit{Self-Efficacy}. For example, P4 elaborated their concerns regarding the app permissions during the interview: \textit{``When you have an app and you see the rights, you have to allow them all... You can't really disable the ones you are not agreeing with.''} P9 from the interview also discussed the insufficiency of using visual summaries such as icons: \textit{``I don't think the icons alone would lead me to take an action... They're just telling you, not really prompting you to do something.''} Participants from the survey similarly commented: \textit{``I just don't really see whether I can agree to something or not, it seems like it is demanded by the app.''} The concerns are also reflected in the participants' suggestions for improving the design. For example, a participant mentioned that they would prefer if \textit{``It could be noted whether the user can disable some of the app's tracking activities.''}

\textbf{Hidden and lack of visual cues} ($N=16$):
Some participants voiced the concern that in the control design, the app privacy information is hidden in the long list of other information about the app. For example, when discussing \textit{Response Cost}, a participant mentioned that \textit{``the average person will not scroll all the way down to read it.''}  A similar concern was voiced by P6 from the interview: \textit{``It doesn't warn you right at the beginning... So you have to really give the app time and be attentive.''} Another participant also mentioned about \textit{Response Efficacy}: \textit{``sometimes we forget to read the smaller words.''} Moreover, participants noted higher \textit{Response Cost} due to limited visual cues to help them digest the information; for example: \textit{``The design does not have any visible and attractive branding for privacy-related emergencies that may occur.''} Participants recommended, e.g., to include\textit{``more colors to differentiate them from other information''} and \textit{``make the more sensitive parts stand out.''}

\subsection{User Feedback on the Treatment Design}
Participants discussed various positive and negative aspects of the treatment design. We first summarize these aspects below; then, we report participants' feedback on each specific design feature.

\subsubsection{Positive factors about the treatment design.}
\hfill

\textbf{The salience of information} ($N=36$):
Participants described different aspects concerning how visible the privacy information is in the treatment design. For example, when discussing \textit{Perceived Vulnerability}, a participant mentioned: \textit{``There are lots of easily seen buttons (Privacy FAQ, Tutorial) that I can see and click to find out more information.''} Some participants also highlighted the use of color-coded texts and icons to convey a sense of risk, which caught their eyes and increased their \textit{Perceived Severity}; e.g., \textit{``Concerning things are in red, which draws my attention and makes me want to find out more;''} and, \textit{``Each part has a symbol to show if there is some risk.''} In the interview, regarding \textit{Response Cost}, P5 also highlighted: \textit{``The [treatment design] is easy as ... the info is right there on the front page in the home screen.''}

\textbf{Navigability} ($N=36$):
The easily navigable information had a positive impact on \textit{Response Efficacy} and \textit{Self Efficacy}, as the following response indicates: \textit{``You only have to click a few buttons to have the information at your disposal, making it easier to take action.''} On a similar note, another participant stated, \textit{``Because the privacy section is well organized, browsing and finding a wanted topic becomes more efficient.''} Some participants pointed out how the navigability of the design influenced \textit{Response Cost}; e.g., \textit{``It was easy to reach my `information privacy', so on that part, yes I do agree this design reduced the effort.''} 

\textbf{Detailed information about privacy} ($N=29$):
Participants found the design comprehensive in explaining the app's data practices, privacy risks, and protective measures. For example, when discussing \textit{Perceived Vulnerability}, a participant stated, \textit{``It does a good job explaining why the app could be intrusive within privacy terms.''} Relating to \textit{Perceived Severity}, participants also highlighted that the design conveys the potential consequences of not taking action to safeguard one's privacy; e.g., \textit{``The reviews and privacy FAQ help me understand how badly my information could be used and how it is all processed.''} Participants acknowledged that the detailed information, particularly through the tutorial, helped them understand the possible actions and increased their perception of \textit{Response Efficacy} and \textit{Self-Efficacy}; e.g., \textit{``Provided detailed steps you can take to protect your data such as stopping marketing ads and sharing only certain things with others;''} and \textit{``There is info letting me know how to take action to attend to my information privacy which makes me confident that I can do so.''} 

\textbf{Facilitate informed decision-making} ($N=25$):
Participants felt that the treatment design guided their privacy-related decisions in the application. This increased \textit{Response Efficacy}, as a participant explained, \textit{``All information provided help the user to make an educated choice.''} Participants also expressed increased \textit{Self Efficacy}; e.g., \textit{``I, as a user, can take to decide if I want to accept the terms, and I can also choose what I want to share.''} In terms of reducing \textit{Response Cost}, the interview participant P9 elaborated on the benefit of having access to the privacy information before downloading the application: \textit{``Instead of having to download the app and then going to the app and checking like `OK, these are the privacy things that you should be aware of', it was already on the interface before I even had to download the app."}

\textbf{Ability to take action} ($N=10$):
Some participants noted how information in the treatment design increased their perception of \textit{Self Efficacy} in attending to their privacy; e.g., \textit{``You only have to click a few buttons to have the information at your disposal, making it easier to take action.''} They also mentioned the ability to control the settings in the app as one important factor; e.g., \textit{``I'm able to change settings about protecting my data while using the app;''} and, \textit{``I can opt-out and choose to be selective of what I share on the app.''}

\subsubsection{Negative factors about the treatment design.} \hfill

\textbf{Information overload} ($N=15$):
While many participants appreciated the amount of detailed information in the design, others felt that the design was too heavy with information that users would not be likely to read. For example, a participant stated, \textit{``If it was an everyday download of the app, I think if I opened the new design [treatment] I would probably not read it...''} The amount of information mainly affected participants' opinion of the \textit{Response Efficacy}, \textit{Self Efficacy}, and \textit{Response Cost}; e.g., \textit{``Too much information, should be less for the user;''} \textit{``Application provides the option, but it seems way too sophisticated for an average user;''} and, \textit{``The long text can be hard to read through.''} Participants recommended further reducing the text and using alternative formats such as graphs or videos to convey the information.

\textbf{Fragmented information}  ($N=7$):
Participants also felt that the privacy information in the design was too scattered. One comment explained, \textit{``Not all available information was easily visible in one place.''} After being introduced to the treatment design, P1 also felt the same during the interview: \textit{``The information is a bit broken apart..., which makes it a bit more difficult to make sure that I have read everything.''} Suggestions proposed by participants included grouping the information to make it easier to identify: \textit{``...I think all things regarding privacy should be kept in one place. I might put it under `Privacy rating', but with red background and here would be all information, FAQs and reviews.''}

\textbf{Discoverability of certain features} ($N=7$):
Some participants were unable to identify the features in the design that were meant to help the user attend to their information privacy. This lowered their opinion of the \textit{Response Efficacy} of the design, as a participant stated: \textit{``I did not see any option that would enable me to attend to the privacy-related problems in this application.''} It also negatively impacted participants' \textit{Self Efficacy}; e.g., \textit{``The privacy-related problems are not addressed on this app. They do not give you an option to not share info.''} A participant also suggested adding \textit{``a search bar to find keywords.''}

\subsubsection{Feedback related to specific design features}
Many participants directly described specific aspects of the five main design features that contributed to their assessment of the application.  

The \textbf{Privacy Rating} was the most well-received design feature, being mentioned in 45 responses. For example, a participant mentioned how the feature affected their \textit{Perceived Vulnerability}: \textit{``The rating received in particular displays severe lack of security with which my data is handled, allowing me to easily note that my privacy is not of major priority to the developer.''} Concerning \textit{Perceived Severity}, a participant explained, \textit{``The rating allows for quickly noticing what egregious errors authors of the app have made.''} A similar reasoning also influenced participants' perception of \textit{Response Efficacy}; e.g., \textit{``The privacy rating is an easy way to figure out if an app is safe or if it has a lot of privacy-related problems.''} While some appreciated the informative red color used for the rating, others desired more prominence; e.g., regarding \textit{Response Costs}, a participant wrote: \textit{``Both `Privacy FAQ' and `Privacy Tutorial'  are easily visible but despite the red color I didn't notice `privacy rating'.''} 

The \textbf{Privacy Reviews} were mentioned in 30 comments, with participants appreciating the input from actual application users; e.g., \textit{``I like the transparency because it showed the privacy reviews from individuals.''} The user reviews helped increase \textit{Perceived Vulnerability} and \textit{Perceived Severity}, as a participant commented, \textit{``Just read the reviews and make my own decision -- if I see that the app has a low score in terms of privacy, I'm immediately on the back foot.''} Some participants felt a greater sense of involvement with the application (and hence \textit{Self Efficacy}) through the privacy review feature; e.g., \textit{``Thanks to this function I can read reviews and also give my own so I feel like I can contribute.''} However, some participants questioned the necessity of separating privacy-specific reviews from regular ones; e.g. \textit{``Privacy `user ratings' feel unnecessary - as anyone can write what they feel anytime in regular ratings.''}.

The information in the \textbf{Privacy FAQ} is referred to in 29 comments. This feature helped participants get a better sense of the anticipated privacy-related problems (i.e., \textit{Perceived Vulnerability} and \textit{Perceived Severity}); e.g., \textit{``In the privacy FAQ we can see how the app manages our data to third party companies or apps that could ask for our data, in that case, we can understand the problems it could bring...''} Some participants viewed the FAQ as a guide to their information privacy, augmenting \textit{Self Efficacy}; e.g., \textit{``Shows you the question and answer section where you can guide yourself.''} P6 from the interview also highlighted that the FAQ was easily accessible: \textit{``It's very informative and you can easily find the frequently asked questions so meaning that you can also maybe probably get help right there.''} Contrastingly, some participants indicated that the FAQ could be further condensed, with a suggestion for \textit{``a TLDR at the start/end of the text.''}

Participants mentioned the \textbf{Privacy Tutorial} in 24 comments. They appreciated its simplicity and succinctness, as evident in this comment: \textit{``The privacy tutorial is clear and easy to understand and straight to the point''}, improving their perceived \textit{Self Efficacy}. They also denoted the support for \textit{Response Efficacy}; e.g., \textit{``In the privacy tutorial we get step by step how to change the information privacy.''} A participant particularly noted that the completeness of the information in the tutorial influenced their \textit{Perceived Vulnerability} and \textit{Perceived Severity}: \textit{``Everything is said on the privacy tutorial. I like that about the app.''} Some other participants, however, suggested a format improvement, for example, to have \textit{``a video in the privacy tutorial with the step-by-step to change our preferences.''}

Although the majority of the participants clicked the ``Get'' button, which triggered the \textbf{Privacy Prompt}, only one participant mentioned this specific prompt in the comments, in relation to their \textit{Perceived Vulnerability}: \textit{``It was very obvious that privacy issues were highlighted; there were prompts to check the security info before getting the app.''} Some participants also suggested incorporating different kinds of prompts, like \textit{``a pop up that says App Store doesn't recommend due to privacy concerns''} or \textit{``a little pop up once in a while to remind users of the privacy information''} within the application.

\section{Discussion}
Our conceptual design reimagines the familiar App Store interface in terms of design elements concerning privacy-related information. Using PMT, we specifically addressed the motivational issue of users' engagement with privacy notices, rather than solely enhancing the representation of the information (e.g.,~\cite{10.1145/2857705.2857741, 9624976}). While some of the design ideas have been approached in previous work, such as privacy rating~\cite{Tsai2011}, icons and visual cues~\cite{10.1145/3411764.3445387, 9624976}, and condensed information~\cite{217470, 10.1145/2857705.2857741}, our work offers a systematic way to consider those ideas in a design space characterized by the PMT components. Results from our user study indicated the intricate relationship between the design ideas targeting different components of PMT and how they collectively impacted users' motivation. Below, we first reflect on the impact of PMT and our design decisions on user motivation, then we discuss the practical implications of our design and the limitations of our study.

\subsection{Design Considerations for Motivating Users to Attend to Privacy}

\subsubsection{Targeting PMT components in combination to induce motivation}
During our design process, we discovered that the PMT components interact with each other to promote user motivation; for the threat appraisal to serve a meaningful purpose, users need to be provided with an effective recourse and mechanism to combat the threat (therefore, an increased \textit{Coping Appraisal}). Similarly, for users to be motivated to undertake a protective action, they need a driving factor, such as a clear understanding of the risks posed by a privacy-related problem (related to the \textit{Threat Appraisal}). Similar trends were observed in previous studies, for example, prior victims of social media harms displayed increased safety protection behavior~\cite{10.1145/3491102.3517643}, individuals informed of card fraud threats were more likely to use secure mobile payments~\cite{255664}, and showing educational videos about security risks of smartphones without screen locks increased the tendency of users to adopt screen locks~\cite{10.5555/3235924.3235929}. Moreover, in our conceptual prototype, even though each design idea was made to address a few specific PMT components, our participants correlated them to other threats and coping mechanisms as well. For instance, the privacy rating was meant to target \textit{Perceived Vulnerability} and \textit{Perceived Severity} by giving the user a sense of the risk associated with the app. But our participants felt that it also offered an easy solution to judge the safety of the application and avoid downloading it, as a participant commented, \textit{``The privacy rating is an easy way to figure out if an app is safe or if it has a lot of privacy-related problems''}, thus influencing their perception of \textit{Response Efficacy} and \textit{Response Costs}. Therefore, an efficient design needs to integrate and address the PMT components as a whole rather than individually. This was also one of the factors that motivated our decision to merge our preliminary design ideas targeting the PMT components separately into one unified design.

\subsubsection{Deciding on the amount and types of information to communicate}
While our design has much less text than the standard privacy policy, some participants still felt it daunting to process all the information in this succinct form. This explains the lower ratings for \textit{Response Cost} for our conceptual design. On the other hand, participants indicated that the information provided by privacy labels in the current App Store lacks sufficient details and actionable insights; as a result, participants did not feel confident to take appropriate actions to attend to their privacy in the application (contributing to low \textit{Self Efficacy}). Thus, deciding on the amount and types of information to communicate is crucial to motivate users to examine privacy-related concerns and also enable them to gain meaningful and actionable insights. For future iterations of our design, we need to further refine the information presentation (and thus reduce \textit{Response Cost}) while maximizing the utility of the information. A perhaps more viable way is to cater to the information needs of different users by providing personalized interfaces.

\subsubsection{Using visual cues to make privacy-related information more prominent} 
During the user study, we noticed that the presentation of the components that we added was very important in highlighting certain PMT components. For example, participants perceived a greater sense of severity and vulnerability on seeing the red color of the privacy rating. Additionally, based on the participants' feedback, the position of the FAQ and privacy tutorial options played an important role in enhancing their \textit{Perceived Severity} and \textit{Perceived Vulnerability}; this design mitigates a common problem that privacy policies are often inconsistently placed~\cite{Proctor2008}. Therefore, augmenting specific design strategies with strong visual cues, such as the use of color and positioning, can be helpful in enhancing the effect of the design ideas derived from PMT.

\subsubsection{Anchoring familiar design ideas to improve user reception and engagement}
We intentionally designed the five main features with elements familiar to users to reduce the learning curve and potential hesitation towards unfamiliar concepts. In our study, the privacy rating received the highest number of mentions, followed by the privacy reviews. A possible explanation is that users were able to identify and relate these features to existing App Store features (i.e., the Ratings and Reviews section). It is also consistent with users' routine to consult the app rating and user reviews before downloading, as indicated by some participants. On the other hand, the privacy prompt was barely mentioned, suggesting that users do not consider it as a separate feature when interacting with the prompt. This resembles common user behavior to ``mechanically'' click the accept or OK button on privacy policy prompts~\cite{7020612}. Overall, our results showed that users tend to engage with elements they have experience with. These patterns can be leveraged to design interfaces that nudge users to visit targeted sections. 

\subsection{Practical Implications of Our Design}

Our study focuses on proposing and validating design concepts rather than suggesting the implementation details of the design ideas. We acknowledge the potential complexity associated with translating these design ideas into complete functionalities. For instance, finding a reliable and standardized method to calculate the privacy rating can be challenging. Our current design integrates one overall privacy rating calculated according to the application's privacy practice. However, assessing the level of risk can be a subjective judgment and can vary from user to user. Techniques using user preference when calculating the privacy rating (e.g.,~\cite{Tsai2011, Cranor2006}) can be considered to overcome this concern. This requires the App Store to integrate mechanisms to explicitly capture the users' privacy profile or preferences. 

For displaying the privacy violation cases in similar apps, there can be a dataset of news articles and reports labeled by app category and further divided into subcategories (e.g., articles related to fitness tracker apps in the category of ``Health and fitness''). Then different strategies can be applied to retrieve the most relevant privacy violation cases for a particular application, such as considering the severity of the incident and the correlation with the privacy practices of the application. However, the task of identifying a suitable authority to oversee the ongoing maintenance and expansion of this dataset is a challenge. Integrating crowdsourcing approaches, such as ToS;DR\footnote{https://tosdr.org}, can also be a valuable alternative.

The current advances in natural language processing (NLP) techniques offer a promising avenue to achieve many proposed features with minimal manual effort, such as identifying the privacy reviews from the pool of user reviews (e.g.,~\cite{zimmeck2016automated}) and generating answers to common privacy questions (e.g.,~\cite{ravichander-etal-2019-question, 217470}). Automated approaches based on NLP may also help address the current practice that requires app developers to self-report their data practices, sometimes leading to misrepresentation~\cite{10.1145/3491102.3502012}. However, the accuracy of the automated techniques requires further examination. Moreover, how to incorporate those techniques in the workflow of privacy information validation needs future investigation.

Additionally, the privacy tutorial informs users of steps they can take to modify the privacy settings or other opt-out options available in the application. However, the coping mechanisms largely depend on the implementation by the developers themselves. The extent and ease of adopting these options may also vary from application to application. The memorability of the presented steps is another limitation of our current design, as there is no easy way for the user to follow the tutorial once the application is launched. Previous work, such as privacy assistants that help the users effectively choose the appropriate settings~\cite{197297, popets} and contextualizing privacy policy during interaction~\cite{10.1145/3491102.3517688}, points to useful directions.  

Finally, we acknowledge Solove's \cite{solove} argument that people cannot be fully engaged in privacy self-management given the large number of applications and services users interact with. Solove differentiated ``privacy'' in a concrete context when people are reasoning the risk of their behavior from the general attitude towards ``privacy''. This distinction is also what we advocate in this work. We hope to motivate users to attend to concrete information related to the privacy practices of a certain application, helping them better reason the risks and their vulnerability and gain self-efficacy. In particular, self-efficacy in PMT directly helps to combat the feeling of futility of ``privacy self-management.'' We also agree with Solove that a bigger impact requires ``regulating the architecture that structures the way information is used, maintained, and transferred''~\cite{solove} so that self-efficacy can sustain. Our current work, however, approaches this from a different angle, motivating users to start to attend to privacy. Such a bottom-up effort (i.e., coming from users) may contribute to accelerating the top-down approach (i.e., coming from regulations) of architectural change. 

\subsection{Limitations and Future Work}
Given how our study was conducted, there are several limiting factors to consider. Firstly, we acknowledge that our design focuses on motivating users to understand privacy concerns when they encounter individual apps rather than comparing several competing apps from the same category. Our work currently also does not convey the common privacy practice of apps from one category. However, the scenario we considered in our study often occurs when users select apps to fulfill their needs, which is an important opportunity for educating users about privacy concerns. The application of our findings to broader app categories or user contexts needs to be explored in future work.

Second, in the user studies, we did not conceal from the participants the purpose of the study (i.e., to investigate design principles that address users' privacy-related motivation). This could have introduced some bias in their interactions with the designs and the type of information they looked for. However, as discussed in Section~\ref{sec:online_study_procedure}, this helped to ensure that the participants were equally exposed to the design features and is aligned with our research objective. Moreover, the phrases and examples used in the conceptual design could have also affected participants' sensitivity to privacy issues and matters. While these phrases and examples were inspired by use cases and possible risks of health apps in real life to provide plausible information to the users, the users' reactions to different tones and phrasing of similar privacy information should be investigated in future work.

Next, the selective recruitment of interview participants based on the quality of their responses may inadvertently favor individuals who are already more engaged or interested in the topic and may overlook the perspectives and behaviors of less motivated users. This may influence the applicability and generalizability of the study's conclusions to the broader user population. In addition, the primary way of measuring the motivation of our participants was through self-reporting based on the survey and the interview. However, the feedback participants gave may not reflect their true motivation and associated further actions. Moreover, motivation changes can occur over time and this change was not captured by our study. A longitudinal field study can be conducted in the future to address this. Lastly, since the study was conducted in English and required a certain level of device exposure, we had to filter out potential participants who did not fit these criteria through Prolific. Future work could focus on privacy-related motivation issues of users who use other languages, from other cultures, or who have different technology proficiencies.

\section{Conclusion}
Awareness of information privacy is crucial for users to navigate today's technology landscape. Past work has focused on improving the presentation of privacy-related information but demonstrated limited effort in systematically addressing mechanisms for enhancing user motivation in privacy awareness and action. To address this gap, we utilized the Protection Motivation Theory (PMT) to inform design ideas aimed at motivating users to increase their interaction with privacy-related information. We put these design ideas into a new conceptual design of the App Store interface and evaluated it in a user study. The results indicated that to effectively promote user motivation, PMT components need to be consolidated and the quantity and types of information communicated in design should be carefully considered, and possibly personalized. Moreover, combining PMT-inspired concepts with other design elements, such as visual cues and user familiarity, can yield favorable outcomes. Our research contributes to design considerations and implications that address the essential motivational issue of augmenting users' privacy awareness and enabling users to take concrete privacy protection actions.

\begin{acks}
We thank our participants for their time and thoughtful answers. We also thank the reviewers for their valuable feedback. This research is supported by the Fonds de recherche du Québec (\grantnum{FRQ}{2022-PR-300101}) and the Canada Research Chairs program (\grantnum{CRC2}{CRC-2021-00076}).
\end{acks}

\bibliographystyle{ACM-Reference-Format}
\bibliography{reference}

\end{document}